\documentclass{iau}
\usepackage{graphicx}
\title[Dust in radio galaxies]{Dust heating in the cores \\
 of 3CRR radio galaxies}
\author[M. Birkinshaw et al.]{M. Birkinshaw,
 D.M. Worrall \and A. Bliss}
\affiliation{
 HH Wills Physics Laboratory, University of Bristol\\
  Tyndall Avenue, Bristol BS8 1TL, U.K.
}
\pubyear{2014}
\volume{313} 
\pagerange{xx--xx}
\jname{Extragalactic jets from every angle}
\editors{F. Massaro, C.C. Cheung, E. Lopez, A. Siemiginowska, eds.}
\begin{document}
\maketitle
\begin{abstract}
 We have undertaken a \textit{Spitzer} campaign to measure the IR
 structures and spectra of low-redshift 3CRR radio galaxies. The
 results show that the $3.6 - 160 \ \rm \mu m$ infrared properties
 vary systematically with integrated source power, and so demonstrate
 that contemporary core activity is characteristic of the
 behaviour of sources over their lifetimes. IR
 synchrotron emission is seen from jets and hotspots in some
 cases. Thermal emission is found from a jet/gas interaction in
 NGC\,7385. Most of the near-IR integrated colours of the low-redshift
 3CRR radio galaxies are similar to those of passive galaxies, so that
 IR colours are poor indicators of radio activity. 
 \keywords{galaxies:active, radio continuum: galaxies, infrared:
  galaxies}
\end{abstract}
\firstsection
\section{Introduction}
We have undertaken an IR imaging campaign for the low-redshift
3CRR \cite[(Laing, Riley \& Longair 1983)]{L83} radio sources using
the IRAC ($3.6$, $4.5$, $5.8$, and $8.0 \ \mu \rm m$ band) and
MIPS ($24$, $70$, and $160 \ \mu \rm m$ band) cameras on the
\textit{Spitzer} satellite 
\cite[(Fazio \etal\ 2004; Rieke \etal\ 2004)]{F04,R04}. Our
main aims were 
(1) to quantify the extent to which signatures of the radio activity
 of the cores appear in the core IR emission; 
(2) to investigate the heating of dust and gas in the core caused by 
 the active nucleus;
(3) to investigate whether the host galaxies display any extended 
 dust features, such as those that might arise from recent
 mergers or from uplifts from the core; and
(4) to image IR emission from radio structures such as jets and hot
 spots, and so to study the relativistic electron populations that
 they contain.

\section{Spitzer imaging}

The sensitivity and angular resolution that the Spitzer cameras
provide is a strong function of wavelength, and the images of the
3CRR galaxies usually show structure in the host galaxy only in the
IRAC bands, where the FWHM of the point spread function (PSF) varies from
$1.6$ to $1.9 \, \rm arcsec$. The $6$ to $40$-arcsec PSFs
of the MIPS images provide structural information only for the closest
objects, such as M\,87 ($=$ 3C\,274), but since the IR emission is usually
concentrated in the cores in the MIPS bands, the uncertainty in the
core flux densities introduced by our inability to make precise fits
to the structures of the galaxies is not usually a limiting factor in
measuring useful core spectra.
 
\begin{figure}[b]
% \vspace*{-2.0 cm}
 \begin{center}
  \includegraphics[width=1.74in]{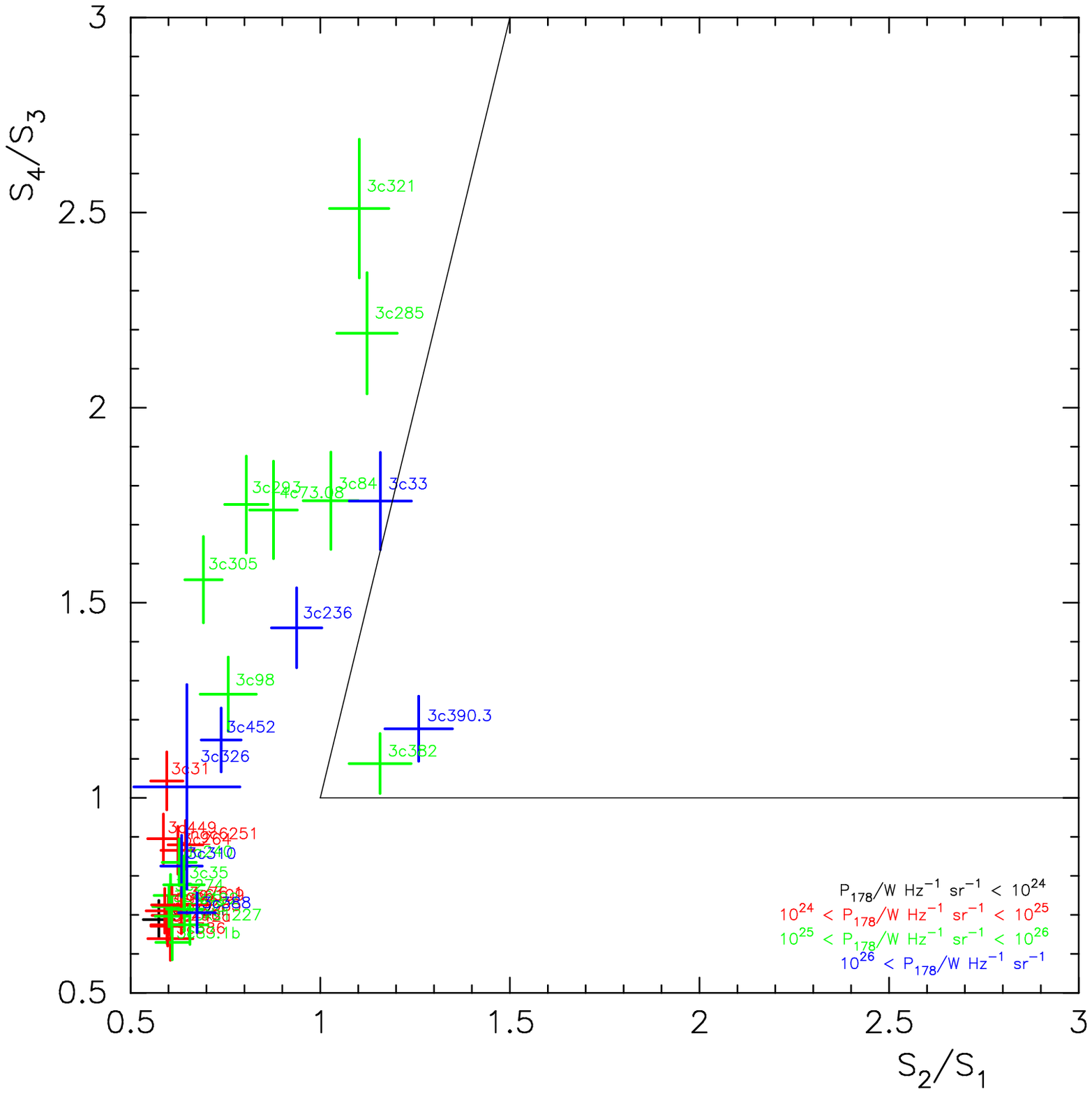} 
  \includegraphics[width=1.74in]{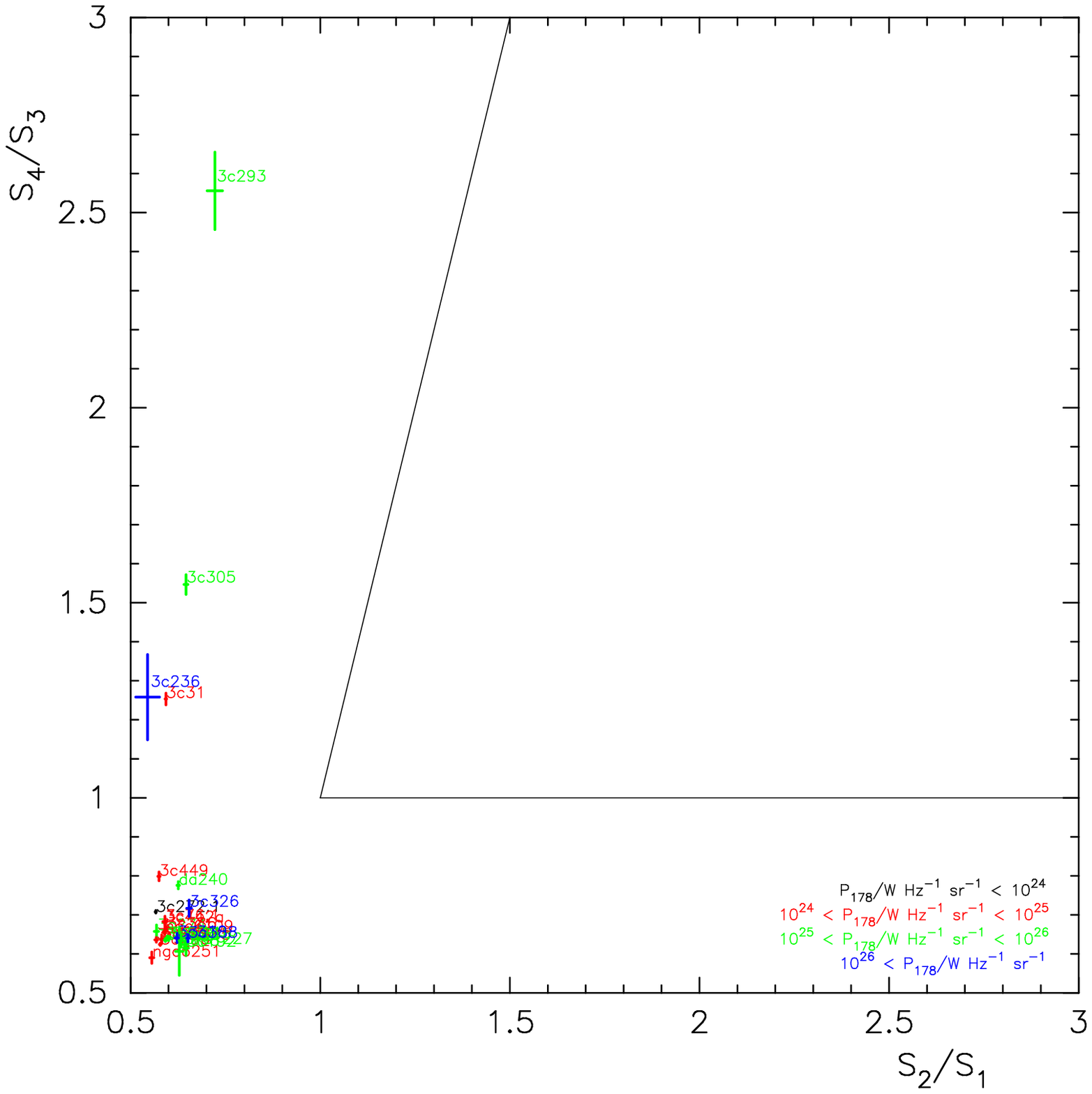} 
  \includegraphics[width=1.74in]{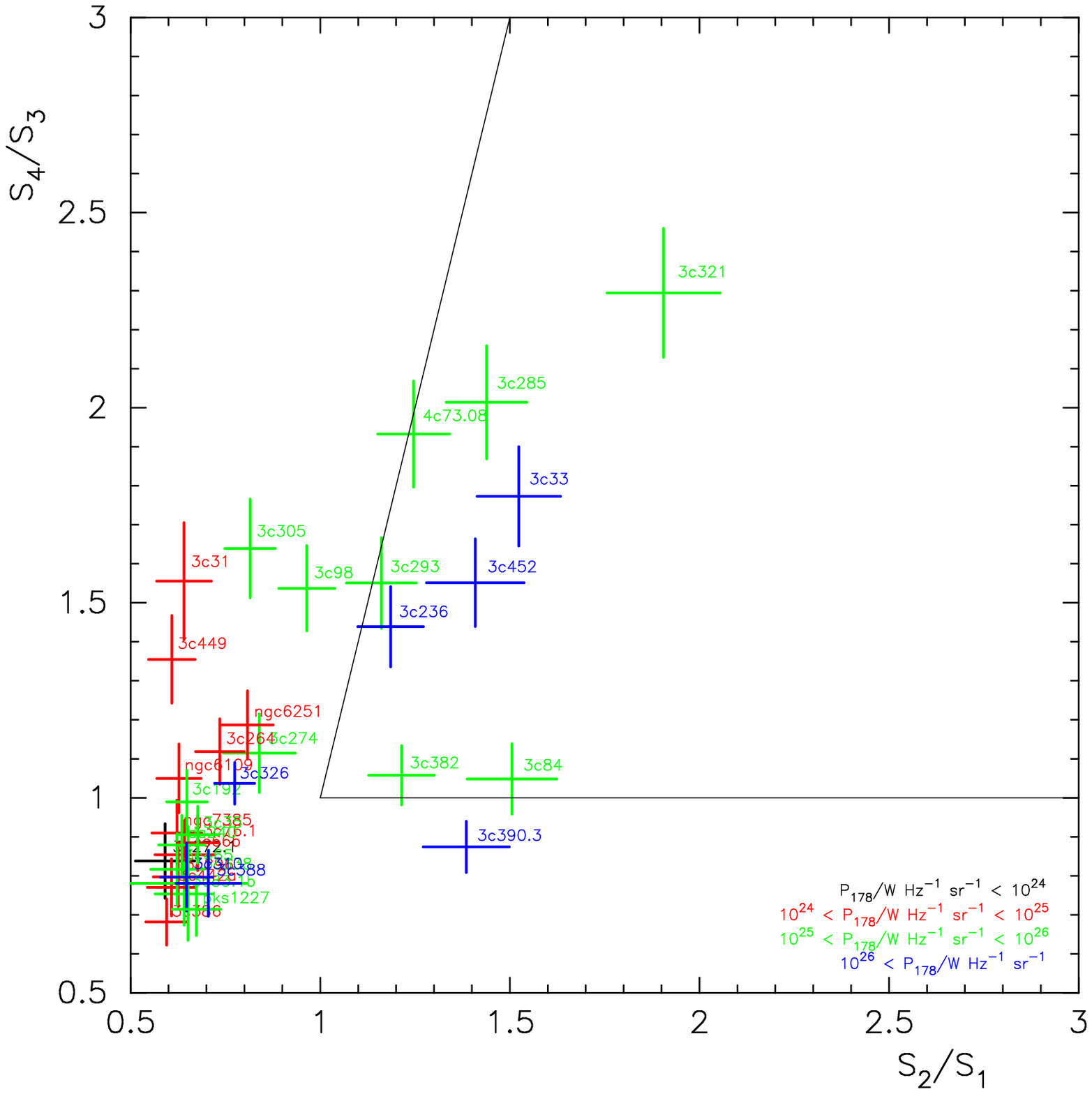} 
% \vspace*{-1.0 cm}
  \caption{IRAC colours of the 3CRR radio galaxies, with the
   $S_4/S_3 \equiv 8.0/5.8 \ \rm \mu m$ colour plotted against the 
   $S_2/S_1 \equiv 4.5/3.6 \ \rm \mu m$ colour.
   Left: measured in a fixed 21-arcsec radius aperture.
   Centre: total light from the host galaxy component, as determined
   from a composite S\'ersic plus point-source model fit. 
   Right: light from the core, excluding host galaxy light to the
   greatest extent possible.
   The lines on each panel separate the region 
   occupied by active galaxies (upper right) from the region 
   occupied by passive galaxies (lower left). These lines were derived
   from the selections of 
   \cite[Stern \etal\ (2005)]{S05} and 
   \cite[Seymour \etal\ (2007)]{S07}.
   Only two 3CRR objects lie in the active-galaxy region in total
   light, and only about 30\% could be identified as active based on
   their core colours. IR surveys are therefore poor at
   finding radio galaxies of low or intermediate power.}
  \label{fig:irac_colours}
 \end{center}
\end{figure}

The basic reduction of the \textit{Spitzer} data used 
\texttt{IRACproc} 
\cite[(Schuster, Marengo \& Patten 2006)]{S06} and MOPEX 
\cite[(Makovoz \& Marleau 2005)]{M05}, and detailed image analysis and
photometry were then performed using codes developed in
Bristol. Photometry for point-like components is accurate to 5\% to
20\%, depending on band, and photometric errors
represented in Figures~\ref{fig:irac_colours}--\ref{fig:ir_spectra}
below include both systematic and statistical components. These
errors become large in the IRAC bands at the high-redshift limit of the
study, $z = 0.1$, because of poor structural discrimination between
the cores and the host galaxies. The errors may also become large for
faint emission in crowded fields, such as in fitting for the IR
emission from the hot spots in 3C\,33 \cite[(Kraft
  \etal\ 2007)]{K07}. 

The IRAC structure of each host galaxy was fitted by an
elliptical S\'ersic model \cite[(S\'ersic 1963)]{S63} with a point-like
component at its centre, convolved with a model PSF, and with
foreground stars, jets, and overlapping galaxy images masked. This
fitting procedure could also be used for the 24-$\mu$m MIPS band in
some cases. For most of the MIPS images, however, a sufficiently good
fit was obtained with a simple point source model because of the
relatively large PSF. The signal/noise of the MIPS flux densities in
the 70- and 160-$\mu$m bands is sometimes quite poor because of the
lumpy background structure in the images.

Some residual images from the fitting showed clear signs of jets, for 
example in 3C\,31 \cite[(Lanz \etal\ 2011)]{L11}, 3C\,66B
\cite[(earlier reported by Tansley \etal\ 2000 from ISO data)]{T00},
and NGC\,6251. In the case of NGC\,6251 the jet is seen to extend well
beyond the optical image of the galaxy, and even beyond the IRAC
frame. When the residual images were compared with the radio
structures other features sometimes became apparent, such as IR-bright
hotspots in cases such as 3C\,33 \cite[(Kraft \etal\ 2007)]{K07}, and
jet interactions, such as the prominent examples in NGC\,7385
\cite[(Rawes \etal\ 2014)]{R14} and 3C\,390.3. 

\section{Spectra}

The colours of the 35 3CRR radio galaxies in the sample are shown in
Fig.~\ref{fig:irac_colours}. The left-hand panel shows the IRAC
$S_4/S_3 \equiv 8.0/5.8 \ \rm \mu m$ colours as a function of the
$S_2/S_1 \equiv 4.5/3.6 \ \rm \mu m$ colours, measured over a fixed
21-arcsec radius aperture. For only two cases do these colours suggest
the presence of an active galaxy, 
according to the criteria of \cite[Stern \etal\ (2005)]{S05} or
\cite[Seymour \etal\ (2007)]{S07}. That is, the average 3CRR 
radio galaxy at low redshift appears passive. The exceptions are the
two broad-line objects in the sample, 3C\,382 and 3C\,390.3. This is
as expected, since the Stern \etal\ and Seymour \etal\ selection cuts
were designed to find broad-line objects. The high $8.0/5.8 \ \rm
\mu m$ ratios evident for some of the galaxies could be interpreted 
as the signature of strong PAH emission rather than an active
nucleus. Many of the higher-power sources in the sample remain in the
region characteristic of pure stellar continuum, which is close to
$(0.7,0.8)$ in this colour-colour plot.

The colours of the host galaxies are shown in the centre panel of
Fig.~\ref{fig:irac_colours}. This clearly shows the passive
nature of the hosts --- the concentration into the region of passive
galaxy colours is more obvious. The most extreme departure from the
locus of passive galaxies occurs for 3C\,293, which is known to be
gas-rich. 

\begin{figure}[b]
% \vspace*{-2.0 cm}
 \begin{center}
  \includegraphics[width=3.5in]{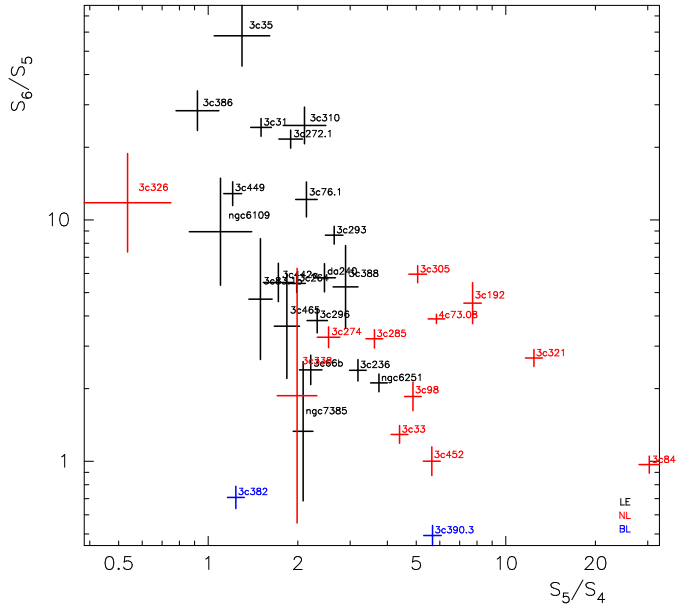} 
% \vspace*{-1.0 cm}
  \caption{Combined MIPS and IRAC colours for the cores of the
   3CRR~radio galaxies. The ratio of $70 \ \rm \mu m$ and $24 \ \rm
   \mu m$ fluxes ($S_6/S_5$) is shown as a function of the ratio of
   $24 \ \rm \mu m$ and $8 \ \rm \mu m$ fluxes ($S_5/S_4$). The
   position of a radio galaxy on this diagram is well-correlated with
   its emission-line type, except that the narrow-line source 3C\,326
   appears anomalous, since it lies in 
   the region occupied by low-excitation objects.}
   \label{fig:irac_mips}
 \end{center}
\end{figure}

When the light from the host galaxies has been removed more of the
3CRR galaxy nuclei show themselves to be active. The core colours are
shown in the right-hand panel of Fig.~\ref{fig:irac_colours}. The
\textit{Spitzer} PSF corresponds to a scale of more than 3~kpc at the
highest redshift in this sample, and so the central stellar emission
in the host galaxy is still a major component of the emission in the 
IRAC bands. This affects the IRAC colours, and the
activity of the nuclei is only apparent for those cores which exhibit
a strong warm-dust emission component. Broad-line objects (such as
3C\,382), narrow-line objects (such as 3C\,285), and low-excitation
objects (such as 3C\,236) can now be found in the Stern \etal\ and
Seymour \etal\ active-galaxy selection region. However, about 70\% of
the 3CRR~radio galaxies would not be selected by near-IR colour, 
even based on their core colours. Improving the
linear resolution for measurement of the IR colours of the cores
would make little difference --- the nearby 3CRR objects for which the
\textit{Spitzer} 
PSF corresponds to sub-kpc linear scales are not notably better
separated from the locus of the passive galaxies than are the more
distant objects.

The line-emission characteristics of the sources are better
distinguished in colours based on a combination of the MIPS and IRAC
bands, as in Fig.~\ref{fig:irac_mips}. The two broad-line emitting
objects lie at the bottom of the diagram, and the low-excitation
sources tend to lie in the upper left, with the narrow-line emitting
cores in the lower right. 3C\,326 is anomalous, with IR colour more
characteristic of a low-excitation source than a narrow-line-emitting
source. This may indicate a misidentification of the core: 3C\,326 has
two compact radio sources near its centre, associated with two
distinct galaxies, with different line-emitting properties. The
optical characteristics of the northern galaxy are more characteristic
of the radio structure \cite[(Rawlings \etal\ 1990)]{R90}, but the
southern galaxy matches the behaviour of other 3CRR~galaxies in its
IR properties.

\begin{figure}[b]
% \vspace*{-2.0 cm}
 \begin{center}
  \includegraphics[height=3.0in]{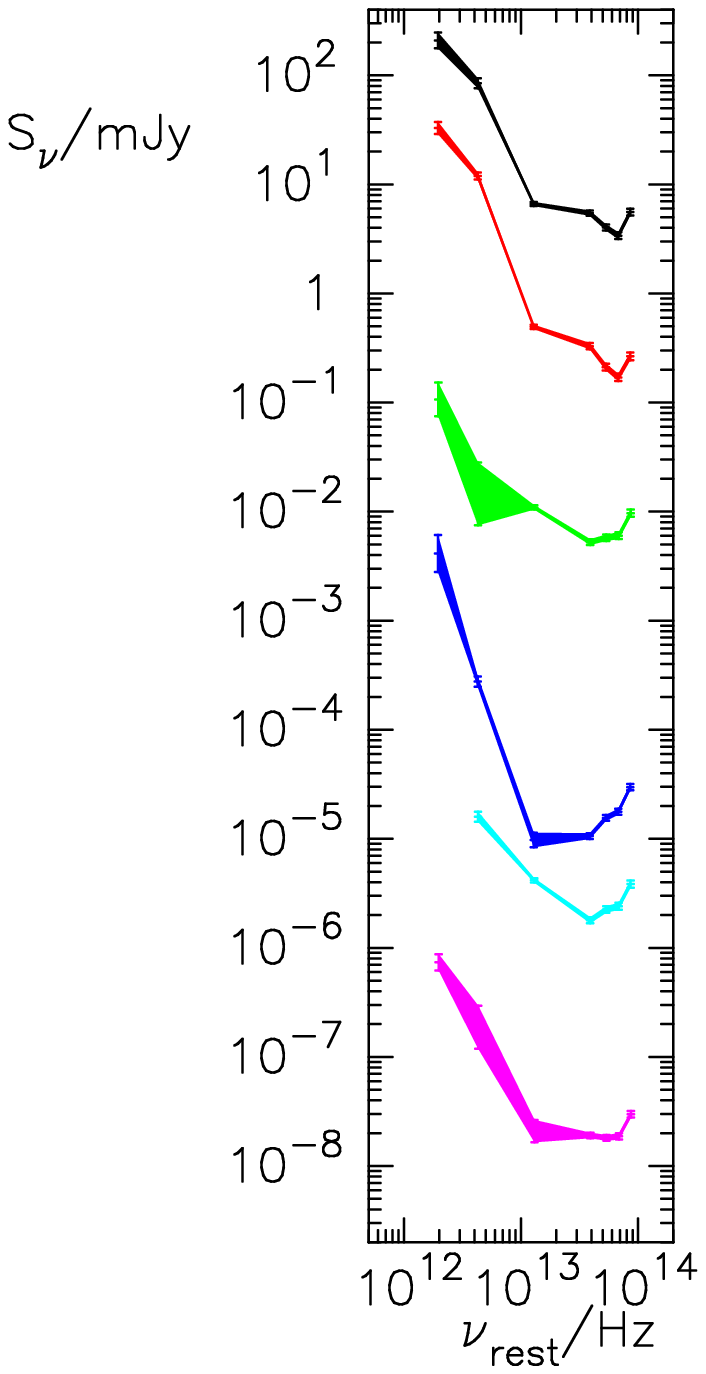} 
%  \hspace{0.3in}
  \includegraphics[height=3.0in]{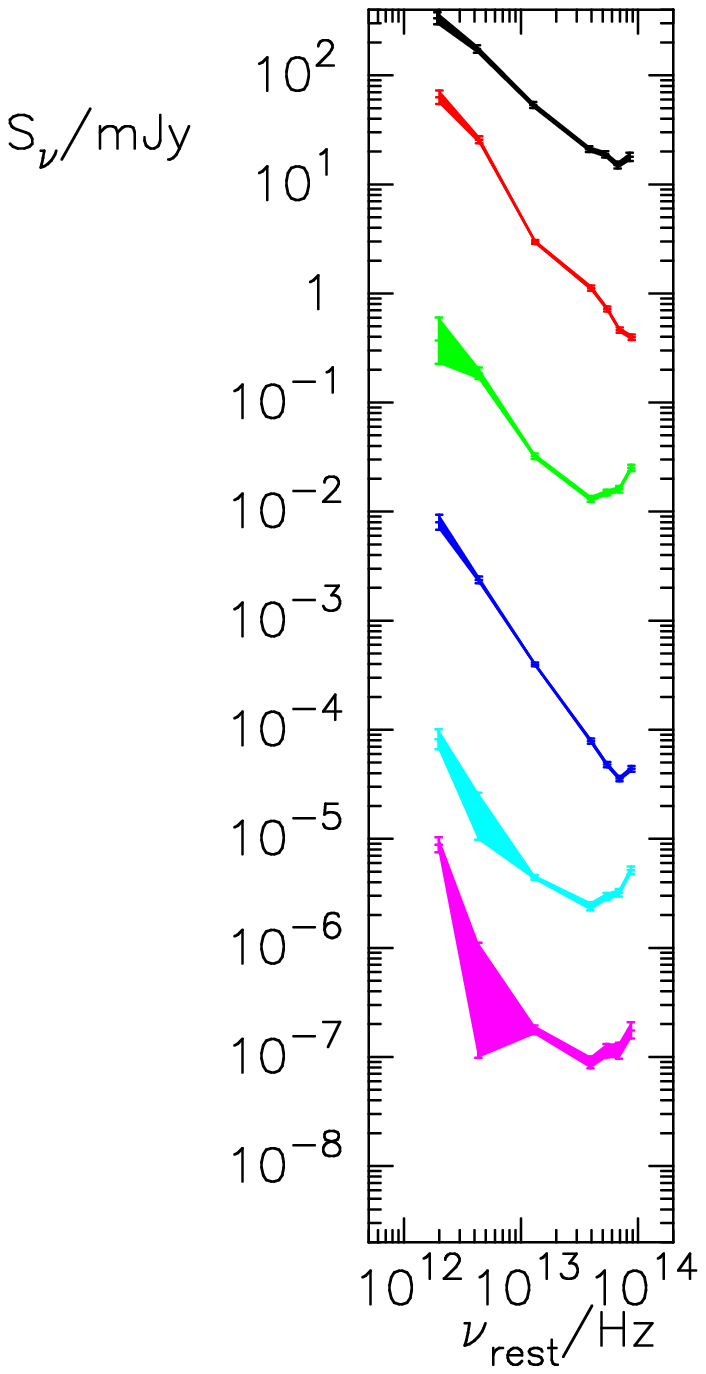} 
%  \hspace{0.3in}
  \includegraphics[height=3.0in]{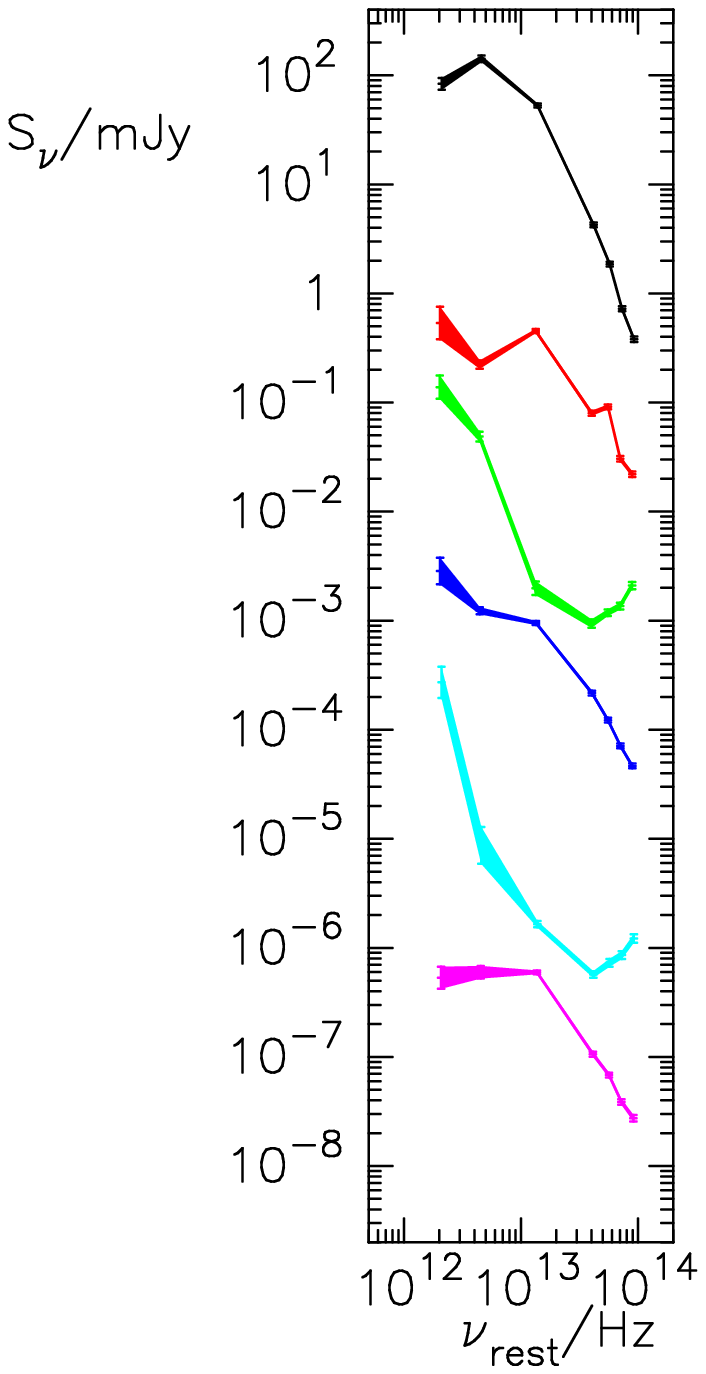} 
% \vspace*{-1.0 cm}
  \caption{IR broad-band fluxes of the cores in representative
    low-power (left), medium-power (centre) and high-power (right)
    3CRR radio galaxies. 178-MHz radio power increases downwards in
    each panel.
    The shape of the IR spectrum of a core is strongly related to
    the total radio power of the 3CRR source.}
  \label{fig:ir_spectra}
 \end{center}
\end{figure}

The distinction between high-power and low-power radio galaxies
becomes clear when the full $3.6 - 160 \ \rm \mu m$ spectra of the
3CRR~galaxies are considered (Fig.~\ref{fig:ir_spectra}). In this
representative set of spectra, the low-power objects (left panel) show
far-IR emission from dust that peaks at lower frequencies than the
high-power objects (right panel) in most cases. Residual stellar
emission is also more evident at low and intermediate radio
powers. The progressive change in shape of the spectra with increasing
radio power strongly suggests the increasing presence of a hot dust
component. The observation that there is a progression of properties
also suggests that, for the majority of the 3CRR~sources, the cores
do not change dramatically on a shorter timescale than the lifetime of
the large-scale radio structure. The lifetimes of the radio sources
exceed 100~Myr, while the heat-loss times for dust near the cores are
far shorter, $< 1$~Myr.

The IR spectra shown in Fig.~\ref{fig:ir_spectra} can be fitted by the
combination of a stellar continuum and dust with a range of
temperatures and/or emissivitities, with the characteristic
temperature of the cool component being near 60~K, and the hotter
component being at about 300~K. There is no evidence for a synchrotron
component in these spectra --- the synchrotron components in 3CRR
sources reported by
\cite[Cleary \etal\ (2007)]{C07} relate to quasars rather than radio
galaxies.

It is notable that there are no strong correlations between the core
radio and IR behaviours, though we find that the core radio flux
density is a slightly better predictor of the core IR flux than is the
total radio flux density.

\section{Conclusions}

There are distinct differences between the core IR spectra,
particularly longward of $24 \ \rm \mu m$, at different source
powers. We believe that this indicates that the current
behaviours of the cores in 3CRR radio galaxies are a reasonable
representation of their average behaviours over the source
lifetimes. However, in detail the relationship is not strong ---
perhaps indicating that the reservoir of material available for future
activity is largely unrelated to the current nuclear activity of the
sources, and that IR emission from the reservoir is a significant
component of the measured fluxes.

The host galaxies show no strong IR peculiarities on scales resolvable
with \textit{Spitzer} in most cases, though some features associated with
radio-source gas interactions have been noted. It is known from HST
imaging that many of these objects show prominent dust lanes
\cite[(e.g., de Koff \etal\ 1996; Martel \etal\ 1999; Sparks
  \etal\ 2000)]{K96,M99,S00}, but this dust probably produces
long-wavelength IR emission that contributes to the 60-K component
that appears in the MIPS camera, and is not separated from the active cores. 

Nevertheless, the angular resolution of \textit{Spitzer} is sufficient
to distinguish radio hot spots, jets, and jet/cloud interactions in a
significant number of cases. Higher angular resolution, and improved
sensitivity, would undoubtedly uncover more examples.

\section{Acknowledgments}

The full team working on the 3CRR low-redshift sample includes the
authors of this short report and P.~Green, H.~Smith, B.~Wilkes,
S.~Willner (CfA), C.~Lawrence (JPL), P.~Barthel (Groningen), E.~Hooper
(Wisconsin), D,~Hines (STScI), and I.~van Bemmel (ASTRON). We thank
M.~Ashby for his advice on processing the IRAC images.
A fuller description of this work is being prepared for publication.

\end{document}